\def\buildarxiv{true}
\documentclass[a4paper,UKenglish,runningheads]{llncs}

\ifdefined\buildself
\usepackage[LNCS,
   key=main,
   year=2021,
   publication={
        Webb BJ., Utting M., Hayes IJ. (2021)
        A Formal Semantics of the GraalVM Intermediate Representation.
        In: Hung D.V., Sokolsky O. (eds)
        Automated Technology for Verification and Analysis. ATVA 2021.},
   orcidicon,
   nocopyright,
   nourl
 ]{authorarchive}
\fi

\ifdefined\buildarxiv
\usepackage{orcidlink}
\renewcommand{\orcidID}[1]{\orcidlink{#1}}
\fi

\listfiles

\usepackage[T1]{fontenc}
\usepackage{lmodern}

\usepackage{tikz}
\usetikzlibrary{arrows.meta}

\usepackage{amsmath}

\usepackage{lib/isabelle,lib/isabellesym}
\usepackage{mathpartir}
\isabellestyle{it}

\newcommand{\showstamps}{0}
\newcommand{\stamp}[1]{\ifthenelse{\equal{\showstamps}{1}}{\alert{#1}}{}}

\usepackage{paper}
\usepackage{alltt}
\usepackage{graphicx} 

\usepackage{caption}
\usepackage{subcaption}
\usepackage{amssymb}
\usepackage{hyperref}

\usepackage{colour-blind}

\Snip{isbinary}{
\begin{center}
\isa{is{\isacharunderscore}{\kern0pt}BinaryArithmeticNode\ {\isacharcolon}{\kern0pt}{\isacharcolon}{\kern0pt}\ IRNode\ {\isasymRightarrow}\ bool}
\end{center}
}%EndSnip
\Snip{inputs_of}{%
\isa{inputs{\isacharunderscore}{\kern0pt}of\ {\isacharcolon}{\kern0pt}{\isacharcolon}{\kern0pt}\ IRNode\ {\isasymRightarrow}\ nat\ list}

\isa{inputs{\isacharunderscore}{\kern0pt}of\ {\isacharparenleft}{\kern0pt}ConstantNode\ const{\isacharparenright}{\kern0pt}\ {\isacharequal}{\kern0pt}\ {\isacharbrackleft}{\kern0pt}{\isacharbrackright}{\kern0pt}}

\isa{inputs{\isacharunderscore}{\kern0pt}of\ {\isacharparenleft}{\kern0pt}ParameterNode\ index{\isacharparenright}{\kern0pt}\ {\isacharequal}{\kern0pt}\ {\isacharbrackleft}{\kern0pt}{\isacharbrackright}{\kern0pt}}

\isa{inputs{\isacharunderscore}{\kern0pt}of\ {\isacharparenleft}{\kern0pt}ValuePhiNode\ nid\ values\ merge{\isacharparenright}{\kern0pt}\ {\isacharequal}{\kern0pt}\ merge\ {\isasymcdot}\ values}

\isa{inputs{\isacharunderscore}{\kern0pt}of\ {\isacharparenleft}{\kern0pt}AddNode\ x\ y{\isacharparenright}{\kern0pt}\ {\isacharequal}{\kern0pt}\ {\isacharbrackleft}{\kern0pt}x{\isacharcomma}{\kern0pt}\ y{\isacharbrackright}{\kern0pt}}

\isa{inputs{\isacharunderscore}{\kern0pt}of\ {\isacharparenleft}{\kern0pt}IfNode\ condition\ trueSuccessor\ falseSuccessor{\isacharparenright}{\kern0pt}\ {\isacharequal}{\kern0pt}\ {\isacharbrackleft}{\kern0pt}condition{\isacharbrackright}{\kern0pt}}
}%EndSnip
\Snip{graphdefnostamp}{
\textbf{typedef} \isa{IRGraph\ {\isacharequal}{\kern0pt}\ {\isacharbraceleft}{\kern0pt}g\ {\isacharcolon}{\kern0pt}{\isacharcolon}{\kern0pt}\ ID\ {\isasymrightharpoonup}\ IRNode\ {\isachardot}{\kern0pt}\ finite\ {\isacharparenleft}{\kern0pt}dom\ g{\isacharparenright}{\kern0pt}{\isacharbraceright}{\kern0pt}}
}%EndSnip
\Snip{helpers_display}{%
\isa{ids\ {\isacharcolon}{\kern0pt}{\isacharcolon}{\kern0pt}\ {\isacharparenleft}{\kern0pt}nat\ {\isasymRightarrow}\ IRNode\ option{\isacharparenright}{\kern0pt}\ {\isasymRightarrow}\ nat\ set}

\isa{ids\ g\ {\isacharequal}{\kern0pt}\ {\isacharbraceleft}{\kern0pt}nid\ {\isasymin}\ dom\ g\ {\isacharbar}{\kern0pt}\ g\ nid\ {\isasymnoteq}\ Some\ NoNode{\isacharbraceright}{\kern0pt}}\\[0.75em]

\isa{kind\ {\isacharcolon}{\kern0pt}{\isacharcolon}{\kern0pt}\ {\isacharparenleft}{\kern0pt}nat\ {\isasymRightarrow}\ IRNode\ option{\isacharparenright}{\kern0pt}\ {\isasymRightarrow}\ nat\ {\isasymRightarrow}\ IRNode}

\isa{kind\ g\ {\isacharequal}{\kern0pt}\ {\isacharparenleft}{\kern0pt}{\isasymlambda}nid{\isachardot}{\kern0pt}\ \textsf{case}\ g\ nid\ \textsf{of}\ None\ {\isasymRightarrow}\ NoNode\ {\isacharbar}{\kern0pt}\ Some\ v\ {\isasymRightarrow}\ v{\isacharparenright}{\kern0pt}}\\[0.75em]

\isa{inputs\ {\isacharcolon}{\kern0pt}{\isacharcolon}{\kern0pt}\ IRGraph\ {\isasymRightarrow}\ nat\ {\isasymRightarrow}\ nat\ set}

\isa{inputs\ g\ nid\ {\isacharequal}{\kern0pt}\ set\ {\isacharparenleft}{\kern0pt}inputs{\isacharunderscore}{\kern0pt}of\ g{\isasymllangle}nid{\isasymrrangle}{\isacharparenright}{\kern0pt}}\\[0.75em]

\isa{succ\ {\isacharcolon}{\kern0pt}{\isacharcolon}{\kern0pt}\ IRGraph\ {\isasymRightarrow}\ nat\ {\isasymRightarrow}\ nat\ set}

\isa{succ\ g\ nid\ {\isacharequal}{\kern0pt}\ set\ {\isacharparenleft}{\kern0pt}successors{\isacharunderscore}{\kern0pt}of\ g{\isasymllangle}nid{\isasymrrangle}{\isacharparenright}{\kern0pt}}\\[0.75em]

\isa{input{\isacharunderscore}{\kern0pt}edges\ {\isacharcolon}{\kern0pt}{\isacharcolon}{\kern0pt}\ IRGraph\ {\isasymRightarrow}\ {\isacharparenleft}{\kern0pt}nat\ {\isasymtimes}\ nat{\isacharparenright}{\kern0pt}\ set}

\isa{input{\isacharunderscore}{\kern0pt}edges\ g\ {\isacharequal}{\kern0pt}\ {\isacharparenleft}{\kern0pt}{\isasymUnion}\isactrlbsub i{\isasymin}ids\ g\isactrlesub \ {\isacharbraceleft}{\kern0pt}{\isacharparenleft}{\kern0pt}i{\isacharcomma}{\kern0pt}\ j{\isacharparenright}{\kern0pt}\ {\isacharbar}{\kern0pt}\ j\ {\isasymin}\ inputs\ g\ i{\isacharbraceright}{\kern0pt}{\isacharparenright}{\kern0pt}}\\[0.75em]

\isa{usages\ {\isacharcolon}{\kern0pt}{\isacharcolon}{\kern0pt}\ IRGraph\ {\isasymRightarrow}\ nat\ {\isasymRightarrow}\ nat\ set}

\isa{usages\ g\ nid\ {\isacharequal}{\kern0pt}\ {\isacharbraceleft}{\kern0pt}j\ {\isasymin}\ ids\ g\ {\isacharbar}{\kern0pt}\ {\isacharparenleft}{\kern0pt}j{\isacharcomma}{\kern0pt}\ nid{\isacharparenright}{\kern0pt}\ {\isasymin}\ input{\isacharunderscore}{\kern0pt}edges\ g{\isacharbraceright}{\kern0pt}}\\[0.75em]

\isa{successor{\isacharunderscore}{\kern0pt}edges\ {\isacharcolon}{\kern0pt}{\isacharcolon}{\kern0pt}\ IRGraph\ {\isasymRightarrow}\ {\isacharparenleft}{\kern0pt}nat\ {\isasymtimes}\ nat{\isacharparenright}{\kern0pt}\ set}

\isa{successor{\isacharunderscore}{\kern0pt}edges\ g\ {\isacharequal}{\kern0pt}\ {\isacharparenleft}{\kern0pt}{\isasymUnion}\isactrlbsub i{\isasymin}ids\ g\isactrlesub \ {\isacharbraceleft}{\kern0pt}{\isacharparenleft}{\kern0pt}i{\isacharcomma}{\kern0pt}\ j{\isacharparenright}{\kern0pt}\ {\isacharbar}{\kern0pt}\ j\ {\isasymin}\ succ\ g\ i{\isacharbraceright}{\kern0pt}{\isacharparenright}{\kern0pt}}\\[0.75em]

\isa{predecessors\ {\isacharcolon}{\kern0pt}{\isacharcolon}{\kern0pt}\ IRGraph\ {\isasymRightarrow}\ nat\ {\isasymRightarrow}\ nat\ set}

\isa{predecessors\ g\ nid\ {\isacharequal}{\kern0pt}\ {\isacharbraceleft}{\kern0pt}j\ {\isasymin}\ ids\ g\ {\isacharbar}{\kern0pt}\ {\isacharparenleft}{\kern0pt}j{\isacharcomma}{\kern0pt}\ nid{\isacharparenright}{\kern0pt}\ {\isasymin}\ successor{\isacharunderscore}{\kern0pt}edges\ g{\isacharbraceright}{\kern0pt}}\\[0.75em]
}%EndSnip
\Snip{wf_start_def}{%
\begin{isabelle}%
wf{\isacharunderscore}{\kern0pt}start\ g\ {\isacharequal}{\kern0pt}\isanewline
{\isacharparenleft}{\kern0pt}{\isadigit{0}}\ {\isasymin}\ ids\ g\ {\isasymand}\ is{\isacharunderscore}{\kern0pt}StartNode\ g{\isasymllangle}{\isadigit{0}}{\isasymrrangle}{\isacharparenright}{\kern0pt}%
\end{isabelle}
}%EndSnip
\Snip{wf_closed_def}{%
\begin{isabelle}%
wf{\isacharunderscore}{\kern0pt}closed\ g\ {\isacharequal}{\kern0pt}\isanewline
{\isacharparenleft}{\kern0pt}{\isasymforall}n{\isasymin}ids\ g{\isachardot}{\kern0pt}\isanewline
\isaindent{{\isacharparenleft}{\kern0pt}\ \ \ }inputs\ g\ n\ {\isasymsubseteq}\ ids\ g\ {\isasymand}\isanewline
\isaindent{{\isacharparenleft}{\kern0pt}\ \ \ }succ\ g\ n\ {\isasymsubseteq}\ ids\ g\ {\isasymand}\ g{\isasymllangle}n{\isasymrrangle}\ {\isasymnoteq}\ NoNode{\isacharparenright}{\kern0pt}%
\end{isabelle}
}%EndSnip
\Snip{wf_phis_def}{%
\begin{isabelle}%
wf{\isacharunderscore}{\kern0pt}phis\ g\ {\isacharequal}{\kern0pt}\isanewline
{\isacharparenleft}{\kern0pt}{\isasymforall}n{\isasymin}ids\ g{\isachardot}{\kern0pt}\isanewline
\isaindent{{\isacharparenleft}{\kern0pt}\ \ \ }is{\isacharunderscore}{\kern0pt}PhiNode\ g{\isasymllangle}n{\isasymrrangle}\ {\isasymlongrightarrow}\isanewline
\isaindent{{\isacharparenleft}{\kern0pt}\ \ \ }{\isacharbar}{\kern0pt}ir{\isacharunderscore}{\kern0pt}values\ g{\isasymllangle}n{\isasymrrangle}{\isacharbar}{\kern0pt}\ {\isacharequal}{\kern0pt}\isanewline
\isaindent{{\isacharparenleft}{\kern0pt}\ \ \ }{\isacharbar}{\kern0pt}ir{\isacharunderscore}{\kern0pt}ends\ g{\isasymllangle}ir{\isacharunderscore}{\kern0pt}merge\ g{\isasymllangle}n{\isasymrrangle}{\isasymrrangle}{\isacharbar}{\kern0pt}{\isacharparenright}{\kern0pt}%
\end{isabelle}
}%EndSnip
\Snip{wf_ends_def}{%
\begin{isabelle}%
wf{\isacharunderscore}{\kern0pt}ends\ g\ {\isacharequal}{\kern0pt}\isanewline
{\isacharparenleft}{\kern0pt}{\isasymforall}n{\isasymin}ids\ g{\isachardot}{\kern0pt}\isanewline
\isaindent{{\isacharparenleft}{\kern0pt}\ \ \ }is{\isacharunderscore}{\kern0pt}AbstractEndNode\ g{\isasymllangle}n{\isasymrrangle}\ {\isasymlongrightarrow}\isanewline
\isaindent{{\isacharparenleft}{\kern0pt}\ \ \ }{\isadigit{0}}\ {\isacharless}{\kern0pt}\ {\isacharbar}{\kern0pt}usages\ g\ n{\isacharbar}{\kern0pt}{\isacharparenright}{\kern0pt}%
\end{isabelle}
}%EndSnip
\Snip{wf_graph_def}{%
\isa{wf{\isacharunderscore}{\kern0pt}graph\ {\isacharcolon}{\kern0pt}{\isacharcolon}{\kern0pt}\ IRGraph\ {\isasymRightarrow}\ bool}

\isa{wf{\isacharunderscore}{\kern0pt}graph\ g\ {\isacharequal}{\kern0pt}\ {\isacharparenleft}{\kern0pt}wf{\isacharunderscore}{\kern0pt}start\ g\ {\isasymand}\ wf{\isacharunderscore}{\kern0pt}closed\ g\ {\isasymand}\ wf{\isacharunderscore}{\kern0pt}phis\ g\ {\isasymand}\ wf{\isacharunderscore}{\kern0pt}ends\ g{\isacharparenright}{\kern0pt}}
}%EndSnip
\Snip{programdef}{%
\textbf{type-synonym} Program = \isa{Signature\ {\isasymrightharpoonup}\ IRGraph}
}%EndSnip
\Snip{heapdef}{%
\isacommand{type{\isacharunderscore}{\kern0pt}synonym}\isamarkupfalse%
\ Heap\ {\isacharequal}{\kern0pt}\ {\isachardoublequoteopen}objref\ {\isasymRightarrow}\ string\ {\isasymRightarrow}\ Value{\isachardoublequoteclose}\isanewline
\isacommand{type{\isacharunderscore}{\kern0pt}synonym}\isamarkupfalse%
\ Free\ {\isacharequal}{\kern0pt}\ {\isachardoublequoteopen}nat{\isachardoublequoteclose}\isanewline
\isacommand{type{\isacharunderscore}{\kern0pt}synonym}\isamarkupfalse%
\ DynamicHeap\ {\isacharequal}{\kern0pt}\ {\isachardoublequoteopen}Heap\ {\isasymtimes}\ Free{\isachardoublequoteclose}%
\\[0.75em]

\isa{h{\isacharunderscore}{\kern0pt}load{\isacharunderscore}{\kern0pt}field\ {\isacharcolon}{\kern0pt}{\isacharcolon}{\kern0pt}} \isa{{\isachardoublequote}{\kern0pt}objref\ {\isasymRightarrow}\ string\ {\isasymRightarrow}\ DynamicHeap\ {\isasymRightarrow}\ Value{\isachardoublequote}{\kern0pt}}

\isa{h{\isacharunderscore}{\kern0pt}load{\isacharunderscore}{\kern0pt}field\ r\ f\ {\isacharparenleft}{\kern0pt}h{\isacharcomma}{\kern0pt}\ n{\isacharparenright}{\kern0pt}\ {\isacharequal}{\kern0pt}\ h\ r\ f}\\[0.75em]

\isa{h{\isacharunderscore}{\kern0pt}store{\isacharunderscore}{\kern0pt}field\ {\isacharcolon}{\kern0pt}{\isacharcolon}{\kern0pt}} \isa{{\isachardoublequote}{\kern0pt}objref\ {\isasymRightarrow}\ string\ {\isasymRightarrow}\ Value\ {\isasymRightarrow}\ DynamicHeap\ {\isasymRightarrow}\ DynamicHeap{\isachardoublequote}{\kern0pt}}

\isa{h{\isacharunderscore}{\kern0pt}store{\isacharunderscore}{\kern0pt}field\ r\ f\ v\ {\isacharparenleft}{\kern0pt}h{\isacharcomma}{\kern0pt}\ n{\isacharparenright}{\kern0pt}\ {\isacharequal}{\kern0pt}\ {\isacharparenleft}{\kern0pt}h{\isacharparenleft}{\kern0pt}r\ {\isacharcolon}{\kern0pt}{\isacharequal}{\kern0pt}\ {\isacharparenleft}{\kern0pt}h\ r{\isacharparenright}{\kern0pt}{\isacharparenleft}{\kern0pt}f\ {\isacharcolon}{\kern0pt}{\isacharequal}{\kern0pt}\ v{\isacharparenright}{\kern0pt}{\isacharparenright}{\kern0pt}{\isacharcomma}{\kern0pt}\ n{\isacharparenright}{\kern0pt}}\\[0.75em]

\isa{h{\isacharunderscore}{\kern0pt}new{\isacharunderscore}{\kern0pt}inst\ {\isacharcolon}{\kern0pt}{\isacharcolon}{\kern0pt}} \isa{{\isachardoublequote}{\kern0pt}DynamicHeap\ {\isasymRightarrow}\ {\isacharparenleft}{\kern0pt}DynamicHeap\ {\isasymtimes}\ Value{\isacharparenright}{\kern0pt}{\isachardoublequote}{\kern0pt}}

\isa{h{\isacharunderscore}{\kern0pt}new{\isacharunderscore}{\kern0pt}inst\ {\isacharparenleft}{\kern0pt}h{\isacharcomma}{\kern0pt}\ n{\isacharparenright}{\kern0pt}\ {\isacharequal}{\kern0pt}\ {\isacharparenleft}{\kern0pt}{\isacharparenleft}{\kern0pt}h{\isacharcomma}{\kern0pt}\ n\ {\isacharplus}{\kern0pt}\ {\isadigit{1}}{\isacharparenright}{\kern0pt}{\isacharcomma}{\kern0pt}\ ObjRef\ {\isacharparenleft}{\kern0pt}Some\ n{\isacharparenright}{\kern0pt}{\isacharparenright}{\kern0pt}}
}%EndSnip
\Snip{ExpressionSemantics}{%
\begin{isamarkuptext}%
\begin{center}
\induct{\isa{{\isacharbrackleft}{\kern0pt}g{\isacharcomma}{\kern0pt}\ m{\isacharcomma}{\kern0pt}\ p{\isacharbrackright}{\kern0pt}\ {\isasymturnstile}\ ConstantNode\ c\ {\isasymmapsto}\ c}}{eval:const}
\induct{\isa{\mbox{}\inferrule{\mbox{{\isacharbrackleft}{\kern0pt}g{\isacharcomma}{\kern0pt}\ m{\isacharcomma}{\kern0pt}\ p{\isacharbrackright}{\kern0pt}\ {\isasymturnstile}\ g{\isasymllangle}x{\isasymrrangle}\ {\isasymmapsto}\ v{\isadigit{1}}}\\\ \mbox{{\isacharbrackleft}{\kern0pt}g{\isacharcomma}{\kern0pt}\ m{\isacharcomma}{\kern0pt}\ p{\isacharbrackright}{\kern0pt}\ {\isasymturnstile}\ g{\isasymllangle}y{\isasymrrangle}\ {\isasymmapsto}\ v{\isadigit{2}}}}{\mbox{{\isacharbrackleft}{\kern0pt}g{\isacharcomma}{\kern0pt}\ m{\isacharcomma}{\kern0pt}\ p{\isacharbrackright}{\kern0pt}\ {\isasymturnstile}\ AddNode\ x\ y\ {\isasymmapsto}\ v{\isadigit{1}}\ {\isacharplus}{\kern0pt}\ v{\isadigit{2}}}}}}{eval:add}
\induct{\isa{{\isacharbrackleft}{\kern0pt}g{\isacharcomma}{\kern0pt}\ m{\isacharcomma}{\kern0pt}\ p{\isacharbrackright}{\kern0pt}\ {\isasymturnstile}\ ParameterNode\ i\ {\isasymmapsto}\ p\ensuremath{_{[\mathit{i}]}}}}{eval:param}
\induct{\isa{{\isacharbrackleft}{\kern0pt}g{\isacharcomma}{\kern0pt}\ m{\isacharcomma}{\kern0pt}\ p{\isacharbrackright}{\kern0pt}\ {\isasymturnstile}\ ValuePhiNode\ nid\ uu\ uv\ {\isasymmapsto}\ m\ nid}}{eval:phi}
\induct{\isa{{\isacharbrackleft}{\kern0pt}g{\isacharcomma}{\kern0pt}\ m{\isacharcomma}{\kern0pt}\ p{\isacharbrackright}{\kern0pt}\ {\isasymturnstile}\ InvokeNode\ nid\ vh\ vi\ vj\ vk\ vl\ {\isasymmapsto}\ m\ nid}}{eval:invoke}
\induct{\isa{{\isacharbrackleft}{\kern0pt}g{\isacharcomma}{\kern0pt}\ m{\isacharcomma}{\kern0pt}\ p{\isacharbrackright}{\kern0pt}\ {\isasymturnstile}\ NewInstanceNode\ nid\ vs\ vt\ vu\ {\isasymmapsto}\ m\ nid}}{eval:invoke}
\induct{\isa{{\isacharbrackleft}{\kern0pt}g{\isacharcomma}{\kern0pt}\ m{\isacharcomma}{\kern0pt}\ p{\isacharbrackright}{\kern0pt}\ {\isasymturnstile}\ LoadFieldNode\ nid\ vv\ vw\ vx\ {\isasymmapsto}\ m\ nid}}{eval:load}
\end{center}%
\end{isamarkuptext}\isamarkuptrue%
}%EndSnip
\Snip{EvalAll}{%
\begin{isamarkuptext}%
\begin{center}
\isa{{\isacharbrackleft}{\kern0pt}g{\isacharcomma}{\kern0pt}\ m{\isacharcomma}{\kern0pt}\ p{\isacharbrackright}{\kern0pt}\ {\isasymturnstile}\ {\isacharbrackleft}{\kern0pt}{\isacharbrackright}{\kern0pt}\ {\isasymlongmapsto}\ {\isacharbrackleft}{\kern0pt}{\isacharbrackright}{\kern0pt}}\\[8px]
\isa{\mbox{}\inferrule{\mbox{{\isacharbrackleft}{\kern0pt}g{\isacharcomma}{\kern0pt}\ m{\isacharcomma}{\kern0pt}\ p{\isacharbrackright}{\kern0pt}\ {\isasymturnstile}\ g{\isasymllangle}nid{\isasymrrangle}\ {\isasymmapsto}\ v}\\\ \mbox{{\isacharbrackleft}{\kern0pt}g{\isacharcomma}{\kern0pt}\ m{\isacharcomma}{\kern0pt}\ p{\isacharbrackright}{\kern0pt}\ {\isasymturnstile}\ xs\ {\isasymlongmapsto}\ vs}}{\mbox{{\isacharbrackleft}{\kern0pt}g{\isacharcomma}{\kern0pt}\ m{\isacharcomma}{\kern0pt}\ p{\isacharbrackright}{\kern0pt}\ {\isasymturnstile}\ nid\ {\isasymcdot}\ xs\ {\isasymlongmapsto}\ v\ {\isasymcdot}\ vs}}}
\end{center}%
\end{isamarkuptext}\isamarkuptrue%
}%EndSnip
\Snip{StepSemantics}{%
\begin{isamarkuptext}%
\begin{center}
\induct{\isa{\mbox{}\inferrule{\mbox{is{\isacharunderscore}{\kern0pt}sequential{\isacharunderscore}{\kern0pt}node\ g{\isasymllangle}nid{\isasymrrangle}}\\\ \mbox{nid{\isacharprime}{\kern0pt}\ {\isacharequal}{\kern0pt}\ {\isacharparenleft}{\kern0pt}successors{\isacharunderscore}{\kern0pt}of\ g{\isasymllangle}nid{\isasymrrangle}{\isacharparenright}{\kern0pt}\ensuremath{_{[\mathit{{\isadigit{0}}}]}}}}{\mbox{{\isacharbrackleft}{\kern0pt}g{\isacharcomma}{\kern0pt}\ p{\isacharbrackright}{\kern0pt}\ {\isasymturnstile}\ {\isacharparenleft}{\kern0pt}nid{\isacharcomma}{\kern0pt}\ m{\isacharcomma}{\kern0pt}\ h{\isacharparenright}{\kern0pt}\ {\isasymrightarrow}\ {\isacharparenleft}{\kern0pt}nid{\isacharprime}{\kern0pt}{\isacharcomma}{\kern0pt}\ m{\isacharcomma}{\kern0pt}\ h{\isacharparenright}{\kern0pt}}}}}{step:seq}
\induct{\isa{\mbox{}\inferrule{\mbox{g{\isasymllangle}nid{\isasymrrangle}\ {\isacharequal}{\kern0pt}\ IfNode\ cond\ tb\ fb}\\\ \mbox{{\isacharbrackleft}{\kern0pt}g{\isacharcomma}{\kern0pt}\ m{\isacharcomma}{\kern0pt}\ p{\isacharbrackright}{\kern0pt}\ {\isasymturnstile}\ g{\isasymllangle}cond{\isasymrrangle}\ {\isasymmapsto}\ val}\\\ \mbox{nid{\isacharprime}{\kern0pt}\ {\isacharequal}{\kern0pt}\ {\isacharparenleft}{\kern0pt}\textsf{if}\ val{\isacharunderscore}{\kern0pt}to{\isacharunderscore}{\kern0pt}bool\ val\ \textsf{then}\ tb\ \textsf{else}\ fb{\isacharparenright}{\kern0pt}}}{\mbox{{\isacharbrackleft}{\kern0pt}g{\isacharcomma}{\kern0pt}\ p{\isacharbrackright}{\kern0pt}\ {\isasymturnstile}\ {\isacharparenleft}{\kern0pt}nid{\isacharcomma}{\kern0pt}\ m{\isacharcomma}{\kern0pt}\ h{\isacharparenright}{\kern0pt}\ {\isasymrightarrow}\ {\isacharparenleft}{\kern0pt}nid{\isacharprime}{\kern0pt}{\isacharcomma}{\kern0pt}\ m{\isacharcomma}{\kern0pt}\ h{\isacharparenright}{\kern0pt}}}}}{step:if}
\induct{\isa{\mbox{}\inferrule{\mbox{is{\isacharunderscore}{\kern0pt}AbstractEndNode\ g{\isasymllangle}nid{\isasymrrangle}}\\\ \mbox{merge\ {\isacharequal}{\kern0pt}\ any{\isacharunderscore}{\kern0pt}usage\ g\ nid}\\\ \mbox{is{\isacharunderscore}{\kern0pt}AbstractMergeNode\ g{\isasymllangle}merge{\isasymrrangle}}\\\ \mbox{i\ {\isacharequal}{\kern0pt}\ find{\isacharunderscore}{\kern0pt}index\ nid\ {\isacharparenleft}{\kern0pt}inputs{\isacharunderscore}{\kern0pt}of\ g{\isasymllangle}merge{\isasymrrangle}{\isacharparenright}{\kern0pt}}\\\ \mbox{phis\ {\isacharequal}{\kern0pt}\ phi{\isacharunderscore}{\kern0pt}list\ g\ merge}\\\ \mbox{inps\ {\isacharequal}{\kern0pt}\ phi{\isacharunderscore}{\kern0pt}inputs\ g\ i\ phis}\\\ \mbox{{\isacharbrackleft}{\kern0pt}g{\isacharcomma}{\kern0pt}\ m{\isacharcomma}{\kern0pt}\ p{\isacharbrackright}{\kern0pt}\ {\isasymturnstile}\ inps\ {\isasymlongmapsto}\ vs}\\\ \mbox{m{\isacharprime}{\kern0pt}\ {\isacharequal}{\kern0pt}\ set{\isacharunderscore}{\kern0pt}phis\ phis\ vs\ m}}{\mbox{{\isacharbrackleft}{\kern0pt}g{\isacharcomma}{\kern0pt}\ p{\isacharbrackright}{\kern0pt}\ {\isasymturnstile}\ {\isacharparenleft}{\kern0pt}nid{\isacharcomma}{\kern0pt}\ m{\isacharcomma}{\kern0pt}\ h{\isacharparenright}{\kern0pt}\ {\isasymrightarrow}\ {\isacharparenleft}{\kern0pt}merge{\isacharcomma}{\kern0pt}\ m{\isacharprime}{\kern0pt}{\isacharcomma}{\kern0pt}\ h{\isacharparenright}{\kern0pt}}}}}{step:end}
\induct{\isa{\mbox{}\inferrule{\mbox{g{\isasymllangle}nid{\isasymrrangle}\ {\isacharequal}{\kern0pt}\ NewInstanceNode\ nid\ class\ frame\ nid{\isacharprime}{\kern0pt}}\\\ \mbox{{\isacharparenleft}{\kern0pt}h{\isacharprime}{\kern0pt}{\isacharcomma}{\kern0pt}\ ref{\isacharparenright}{\kern0pt}\ {\isacharequal}{\kern0pt}\ h{\isacharunderscore}{\kern0pt}new{\isacharunderscore}{\kern0pt}inst\ h}\\\ \mbox{m{\isacharprime}{\kern0pt}\ {\isacharequal}{\kern0pt}\ m{\isacharparenleft}{\kern0pt}nid\ {\isacharcolon}{\kern0pt}{\isacharequal}{\kern0pt}\ ref{\isacharparenright}{\kern0pt}}}{\mbox{{\isacharbrackleft}{\kern0pt}g{\isacharcomma}{\kern0pt}\ p{\isacharbrackright}{\kern0pt}\ {\isasymturnstile}\ {\isacharparenleft}{\kern0pt}nid{\isacharcomma}{\kern0pt}\ m{\isacharcomma}{\kern0pt}\ h{\isacharparenright}{\kern0pt}\ {\isasymrightarrow}\ {\isacharparenleft}{\kern0pt}nid{\isacharprime}{\kern0pt}{\isacharcomma}{\kern0pt}\ m{\isacharprime}{\kern0pt}{\isacharcomma}{\kern0pt}\ h{\isacharprime}{\kern0pt}{\isacharparenright}{\kern0pt}}}}}{step:newinst}
\induct{\isa{\mbox{}\inferrule{\mbox{g{\isasymllangle}nid{\isasymrrangle}\ {\isacharequal}{\kern0pt}\ LoadFieldNode\ nid\ f\ {\isacharparenleft}{\kern0pt}Some\ obj{\isacharparenright}{\kern0pt}\ nid{\isacharprime}{\kern0pt}}\\\ \mbox{{\isacharbrackleft}{\kern0pt}g{\isacharcomma}{\kern0pt}\ m{\isacharcomma}{\kern0pt}\ p{\isacharbrackright}{\kern0pt}\ {\isasymturnstile}\ g{\isasymllangle}obj{\isasymrrangle}\ {\isasymmapsto}\ ObjRef\ ref}\\\ \mbox{h{\isacharunderscore}{\kern0pt}load{\isacharunderscore}{\kern0pt}field\ ref\ f\ h\ {\isacharequal}{\kern0pt}\ v}\\\ \mbox{m{\isacharprime}{\kern0pt}\ {\isacharequal}{\kern0pt}\ m{\isacharparenleft}{\kern0pt}nid\ {\isacharcolon}{\kern0pt}{\isacharequal}{\kern0pt}\ v{\isacharparenright}{\kern0pt}}}{\mbox{{\isacharbrackleft}{\kern0pt}g{\isacharcomma}{\kern0pt}\ p{\isacharbrackright}{\kern0pt}\ {\isasymturnstile}\ {\isacharparenleft}{\kern0pt}nid{\isacharcomma}{\kern0pt}\ m{\isacharcomma}{\kern0pt}\ h{\isacharparenright}{\kern0pt}\ {\isasymrightarrow}\ {\isacharparenleft}{\kern0pt}nid{\isacharprime}{\kern0pt}{\isacharcomma}{\kern0pt}\ m{\isacharprime}{\kern0pt}{\isacharcomma}{\kern0pt}\ h{\isacharparenright}{\kern0pt}}}}}{step:load}
\induct{\isa{\mbox{}\inferrule{\mbox{g{\isasymllangle}nid{\isasymrrangle}\ {\isacharequal}{\kern0pt}\ StoreFieldNode\ nid\ f\ newval\ uu\ {\isacharparenleft}{\kern0pt}Some\ obj{\isacharparenright}{\kern0pt}\ nid{\isacharprime}{\kern0pt}}\\\ \mbox{{\isacharbrackleft}{\kern0pt}g{\isacharcomma}{\kern0pt}\ m{\isacharcomma}{\kern0pt}\ p{\isacharbrackright}{\kern0pt}\ {\isasymturnstile}\ g{\isasymllangle}newval{\isasymrrangle}\ {\isasymmapsto}\ val}\\\ \mbox{{\isacharbrackleft}{\kern0pt}g{\isacharcomma}{\kern0pt}\ m{\isacharcomma}{\kern0pt}\ p{\isacharbrackright}{\kern0pt}\ {\isasymturnstile}\ g{\isasymllangle}obj{\isasymrrangle}\ {\isasymmapsto}\ ObjRef\ ref}\\\ \mbox{h{\isacharprime}{\kern0pt}\ {\isacharequal}{\kern0pt}\ h{\isacharunderscore}{\kern0pt}store{\isacharunderscore}{\kern0pt}field\ ref\ f\ val\ h}\\\ \mbox{m{\isacharprime}{\kern0pt}\ {\isacharequal}{\kern0pt}\ m{\isacharparenleft}{\kern0pt}nid\ {\isacharcolon}{\kern0pt}{\isacharequal}{\kern0pt}\ val{\isacharparenright}{\kern0pt}}}{\mbox{{\isacharbrackleft}{\kern0pt}g{\isacharcomma}{\kern0pt}\ p{\isacharbrackright}{\kern0pt}\ {\isasymturnstile}\ {\isacharparenleft}{\kern0pt}nid{\isacharcomma}{\kern0pt}\ m{\isacharcomma}{\kern0pt}\ h{\isacharparenright}{\kern0pt}\ {\isasymrightarrow}\ {\isacharparenleft}{\kern0pt}nid{\isacharprime}{\kern0pt}{\isacharcomma}{\kern0pt}\ m{\isacharprime}{\kern0pt}{\isacharcomma}{\kern0pt}\ h{\isacharprime}{\kern0pt}{\isacharparenright}{\kern0pt}}}}}{step:store}
\end{center}%
\end{isamarkuptext}\isamarkuptrue%
}%EndSnip
\Snip{TopStepSemantics}{%
\begin{isamarkuptext}%
\begin{center}
\induct{\isa{\mbox{}\inferrule{\mbox{{\isacharbrackleft}{\kern0pt}g{\isacharcomma}{\kern0pt}\ p{\isacharbrackright}{\kern0pt}\ {\isasymturnstile}\ {\isacharparenleft}{\kern0pt}nid{\isacharcomma}{\kern0pt}\ m{\isacharcomma}{\kern0pt}\ h{\isacharparenright}{\kern0pt}\ {\isasymrightarrow}\ {\isacharparenleft}{\kern0pt}nid{\isacharprime}{\kern0pt}{\isacharcomma}{\kern0pt}\ m{\isacharprime}{\kern0pt}{\isacharcomma}{\kern0pt}\ h{\isacharprime}{\kern0pt}{\isacharparenright}{\kern0pt}}}{\mbox{P\ {\isasymturnstile}\ {\isacharparenleft}{\kern0pt}{\isacharparenleft}{\kern0pt}g{\isacharcomma}{\kern0pt}\ nid{\isacharcomma}{\kern0pt}\ m{\isacharcomma}{\kern0pt}\ p{\isacharparenright}{\kern0pt}\ {\isasymcdot}\ stk{\isacharcomma}{\kern0pt}\ h{\isacharparenright}{\kern0pt}\ {\isasymlongrightarrow}\ {\isacharparenleft}{\kern0pt}{\isacharparenleft}{\kern0pt}g{\isacharcomma}{\kern0pt}\ nid{\isacharprime}{\kern0pt}{\isacharcomma}{\kern0pt}\ m{\isacharprime}{\kern0pt}{\isacharcomma}{\kern0pt}\ p{\isacharparenright}{\kern0pt}\ {\isasymcdot}\ stk{\isacharcomma}{\kern0pt}\ h{\isacharprime}{\kern0pt}{\isacharparenright}{\kern0pt}}}}}{top:lift}
\induct{\isa{\mbox{}\inferrule{\mbox{is{\isacharunderscore}{\kern0pt}Invoke\ g{\isasymllangle}nid{\isasymrrangle}}\\\ \mbox{callTarget\ {\isacharequal}{\kern0pt}\ ir{\isacharunderscore}{\kern0pt}callTarget\ g{\isasymllangle}nid{\isasymrrangle}}\\\ \mbox{g{\isasymllangle}callTarget{\isasymrrangle}\ {\isacharequal}{\kern0pt}\ MethodCallTargetNode\ targetMethod\ arguments}\\\ \mbox{Some\ targetGraph\ {\isacharequal}{\kern0pt}\ P\ targetMethod}\\\ \mbox{m{\isacharprime}{\kern0pt}\ {\isacharequal}{\kern0pt}\ new{\isacharunderscore}{\kern0pt}map{\isacharunderscore}{\kern0pt}state}\\\ \mbox{{\isacharbrackleft}{\kern0pt}g{\isacharcomma}{\kern0pt}\ m{\isacharcomma}{\kern0pt}\ p{\isacharbrackright}{\kern0pt}\ {\isasymturnstile}\ arguments\ {\isasymlongmapsto}\ p{\isacharprime}{\kern0pt}}}{\mbox{P\ {\isasymturnstile}\ {\isacharparenleft}{\kern0pt}{\isacharparenleft}{\kern0pt}g{\isacharcomma}{\kern0pt}\ nid{\isacharcomma}{\kern0pt}\ m{\isacharcomma}{\kern0pt}\ p{\isacharparenright}{\kern0pt}\ {\isasymcdot}\ stk{\isacharcomma}{\kern0pt}\ h{\isacharparenright}{\kern0pt}\ {\isasymlongrightarrow}\ {\isacharparenleft}{\kern0pt}{\isacharparenleft}{\kern0pt}targetGraph{\isacharcomma}{\kern0pt}\ {\isadigit{0}}{\isacharcomma}{\kern0pt}\ m{\isacharprime}{\kern0pt}{\isacharcomma}{\kern0pt}\ p{\isacharprime}{\kern0pt}{\isacharparenright}{\kern0pt}\ {\isasymcdot}\ {\isacharparenleft}{\kern0pt}g{\isacharcomma}{\kern0pt}\ nid{\isacharcomma}{\kern0pt}\ m{\isacharcomma}{\kern0pt}\ p{\isacharparenright}{\kern0pt}\ {\isasymcdot}\ stk{\isacharcomma}{\kern0pt}\ h{\isacharparenright}{\kern0pt}}}}}{top:invoke}
\induct{\isa{\mbox{}\inferrule{\mbox{g{\isasymllangle}nid{\isasymrrangle}\ {\isacharequal}{\kern0pt}\ ReturnNode\ {\isacharparenleft}{\kern0pt}Some\ expr{\isacharparenright}{\kern0pt}\ uu}\\\ \mbox{{\isacharbrackleft}{\kern0pt}g{\isacharcomma}{\kern0pt}\ m{\isacharcomma}{\kern0pt}\ p{\isacharbrackright}{\kern0pt}\ {\isasymturnstile}\ g{\isasymllangle}expr{\isasymrrangle}\ {\isasymmapsto}\ v}\\\ \mbox{cm{\isacharprime}{\kern0pt}\ {\isacharequal}{\kern0pt}\ cm{\isacharparenleft}{\kern0pt}cnid\ {\isacharcolon}{\kern0pt}{\isacharequal}{\kern0pt}\ v{\isacharparenright}{\kern0pt}}\\\ \mbox{cnid{\isacharprime}{\kern0pt}\ {\isacharequal}{\kern0pt}\ {\isacharparenleft}{\kern0pt}successors{\isacharunderscore}{\kern0pt}of\ cg{\isasymllangle}cnid{\isasymrrangle}{\isacharparenright}{\kern0pt}\ensuremath{_{[\mathit{{\isadigit{0}}}]}}}}{\mbox{P\ {\isasymturnstile}\ {\isacharparenleft}{\kern0pt}{\isacharparenleft}{\kern0pt}g{\isacharcomma}{\kern0pt}\ nid{\isacharcomma}{\kern0pt}\ m{\isacharcomma}{\kern0pt}\ p{\isacharparenright}{\kern0pt}\ {\isasymcdot}\ {\isacharparenleft}{\kern0pt}cg{\isacharcomma}{\kern0pt}\ cnid{\isacharcomma}{\kern0pt}\ cm{\isacharcomma}{\kern0pt}\ cp{\isacharparenright}{\kern0pt}\ {\isasymcdot}\ stk{\isacharcomma}{\kern0pt}\ h{\isacharparenright}{\kern0pt}\ {\isasymlongrightarrow}\ {\isacharparenleft}{\kern0pt}{\isacharparenleft}{\kern0pt}cg{\isacharcomma}{\kern0pt}\ cnid{\isacharprime}{\kern0pt}{\isacharcomma}{\kern0pt}\ cm{\isacharprime}{\kern0pt}{\isacharcomma}{\kern0pt}\ cp{\isacharparenright}{\kern0pt}\ {\isasymcdot}\ stk{\isacharcomma}{\kern0pt}\ h{\isacharparenright}{\kern0pt}}}}}{top:return}
\induct{\isa{\mbox{}\inferrule{\mbox{g{\isasymllangle}nid{\isasymrrangle}\ {\isacharequal}{\kern0pt}\ UnwindNode\ exception}\\\ \mbox{{\isacharbrackleft}{\kern0pt}g{\isacharcomma}{\kern0pt}\ m{\isacharcomma}{\kern0pt}\ p{\isacharbrackright}{\kern0pt}\ {\isasymturnstile}\ g{\isasymllangle}exception{\isasymrrangle}\ {\isasymmapsto}\ e}\\\ \mbox{cg{\isasymllangle}cnid{\isasymrrangle}\ {\isacharequal}{\kern0pt}\ InvokeWithExceptionNode\ uw\ ux\ uy\ uz\ va\ vb\ exEdge}\\\ \mbox{cm{\isacharprime}{\kern0pt}\ {\isacharequal}{\kern0pt}\ cm{\isacharparenleft}{\kern0pt}cnid\ {\isacharcolon}{\kern0pt}{\isacharequal}{\kern0pt}\ e{\isacharparenright}{\kern0pt}}}{\mbox{P\ {\isasymturnstile}\ {\isacharparenleft}{\kern0pt}{\isacharparenleft}{\kern0pt}g{\isacharcomma}{\kern0pt}\ nid{\isacharcomma}{\kern0pt}\ m{\isacharcomma}{\kern0pt}\ p{\isacharparenright}{\kern0pt}\ {\isasymcdot}\ {\isacharparenleft}{\kern0pt}cg{\isacharcomma}{\kern0pt}\ cnid{\isacharcomma}{\kern0pt}\ cm{\isacharcomma}{\kern0pt}\ cp{\isacharparenright}{\kern0pt}\ {\isasymcdot}\ stk{\isacharcomma}{\kern0pt}\ h{\isacharparenright}{\kern0pt}\ {\isasymlongrightarrow}\ {\isacharparenleft}{\kern0pt}{\isacharparenleft}{\kern0pt}cg{\isacharcomma}{\kern0pt}\ exEdge{\isacharcomma}{\kern0pt}\ cm{\isacharprime}{\kern0pt}{\isacharcomma}{\kern0pt}\ cp{\isacharparenright}{\kern0pt}\ {\isasymcdot}\ stk{\isacharcomma}{\kern0pt}\ h{\isacharparenright}{\kern0pt}}}}}{top:unwind}
\end{center}%
\end{isamarkuptext}\isamarkuptrue%
}%EndSnip
\Snip{AddNodeRules}{%
\begin{center}
\isa{\mbox{}\inferrule{\mbox{g{\isasymllangle}x{\isasymrrangle}\ {\isacharequal}{\kern0pt}\ ConstantNode\ c{\isacharunderscore}{\kern0pt}{\isadigit{1}}}\\\ \mbox{g{\isasymllangle}y{\isasymrrangle}\ {\isacharequal}{\kern0pt}\ ConstantNode\ c{\isacharunderscore}{\kern0pt}{\isadigit{2}}}\\\ \mbox{val\ {\isacharequal}{\kern0pt}\ intval{\isacharunderscore}{\kern0pt}add\ c{\isacharunderscore}{\kern0pt}{\isadigit{1}}\ c{\isacharunderscore}{\kern0pt}{\isadigit{2}}}}{\mbox{CanonicalizeAdd\ g\ {\isacharparenleft}{\kern0pt}AddNode\ x\ y{\isacharparenright}{\kern0pt}\ {\isacharparenleft}{\kern0pt}ConstantNode\ val{\isacharparenright}{\kern0pt}}}}\\[8px]
\isa{\mbox{}\inferrule{\mbox{g{\isasymllangle}x{\isasymrrangle}\ {\isacharequal}{\kern0pt}\ ConstantNode\ c{\isacharunderscore}{\kern0pt}{\isadigit{1}}}\\\ \mbox{{\isasymnot}\ is{\isacharunderscore}{\kern0pt}ConstantNode\ g{\isasymllangle}y{\isasymrrangle}}\\\ \mbox{c{\isacharunderscore}{\kern0pt}{\isadigit{1}}\ {\isacharequal}{\kern0pt}\ IntVal{\isadigit{3}}{\isadigit{2}}\ {\isadigit{0}}}}{\mbox{CanonicalizeAdd\ g\ {\isacharparenleft}{\kern0pt}AddNode\ x\ y{\isacharparenright}{\kern0pt}\ {\isacharparenleft}{\kern0pt}RefNode\ y{\isacharparenright}{\kern0pt}}}}\\[8px]
\isa{\mbox{}\inferrule{\mbox{{\isasymnot}\ is{\isacharunderscore}{\kern0pt}ConstantNode\ g{\isasymllangle}x{\isasymrrangle}}\\\ \mbox{g{\isasymllangle}y{\isasymrrangle}\ {\isacharequal}{\kern0pt}\ ConstantNode\ c{\isacharunderscore}{\kern0pt}{\isadigit{2}}}\\\ \mbox{c{\isacharunderscore}{\kern0pt}{\isadigit{2}}\ {\isacharequal}{\kern0pt}\ IntVal{\isadigit{3}}{\isadigit{2}}\ {\isadigit{0}}}}{\mbox{CanonicalizeAdd\ g\ {\isacharparenleft}{\kern0pt}AddNode\ x\ y{\isacharparenright}{\kern0pt}\ {\isacharparenleft}{\kern0pt}RefNode\ x{\isacharparenright}{\kern0pt}}}}
\end{center}
}%EndSnip
\Snip{AddNodeProof}{%
\isa{{\isasymlbrakk}CanonicalizeAdd\ g\ before\ after{\isacharsemicolon}{\kern0pt}\ wf{\isacharunderscore}{\kern0pt}graph\ g\ {\isasymand}\ wf{\isacharunderscore}{\kern0pt}stamps\ g{\isacharsemicolon}{\kern0pt}\ {\isacharbrackleft}{\kern0pt}g{\isacharcomma}{\kern0pt}\ m{\isacharcomma}{\kern0pt}\ p{\isacharbrackright}{\kern0pt}\ {\isasymturnstile}\ before\ {\isasymmapsto}\ IntVal{\isadigit{3}}{\isadigit{2}}\ res{\isacharsemicolon}{\kern0pt}\ {\isacharbrackleft}{\kern0pt}g{\isacharcomma}{\kern0pt}\ m{\isacharcomma}{\kern0pt}\ p{\isacharbrackright}{\kern0pt}\ {\isasymturnstile}\ after\ {\isasymmapsto}\ IntVal{\isadigit{3}}{\isadigit{2}}\ res{\isacharprime}{\kern0pt}{\isasymrbrakk}\ {\isasymLongrightarrow}\ res\ {\isacharequal}{\kern0pt}\ res{\isacharprime}{\kern0pt}}
}%EndSnip
\Snip{Stutter}{%
\begin{isamarkuptext}%
\begin{center}
\isa{\mbox{}\inferrule{\mbox{{\isacharbrackleft}{\kern0pt}g{\isacharcomma}{\kern0pt}\ p{\isacharbrackright}{\kern0pt}\ {\isasymturnstile}\ {\isacharparenleft}{\kern0pt}nid{\isacharcomma}{\kern0pt}\ m{\isacharcomma}{\kern0pt}\ h{\isacharparenright}{\kern0pt}\ {\isasymrightarrow}\ {\isacharparenleft}{\kern0pt}nid{\isacharprime}{\kern0pt}{\isacharcomma}{\kern0pt}\ m{\isacharcomma}{\kern0pt}\ h{\isacharparenright}{\kern0pt}}}{\mbox{{\isacharbrackleft}{\kern0pt}g{\isacharcomma}{\kern0pt}\ m{\isacharcomma}{\kern0pt}\ p{\isacharcomma}{\kern0pt}\ h{\isacharbrackright}{\kern0pt}\ {\isasymturnstile}\ nid\ {\isasymleadsto}\ nid{\isacharprime}{\kern0pt}}}}\\[8px]
\isa{\mbox{}\inferrule{\mbox{{\isacharbrackleft}{\kern0pt}g{\isacharcomma}{\kern0pt}\ p{\isacharbrackright}{\kern0pt}\ {\isasymturnstile}\ {\isacharparenleft}{\kern0pt}nid{\isacharcomma}{\kern0pt}\ m{\isacharcomma}{\kern0pt}\ h{\isacharparenright}{\kern0pt}\ {\isasymrightarrow}\ {\isacharparenleft}{\kern0pt}nid{\isacharprime}{\kern0pt}{\isacharprime}{\kern0pt}{\isacharcomma}{\kern0pt}\ m{\isacharcomma}{\kern0pt}\ h{\isacharparenright}{\kern0pt}}\\\ \mbox{{\isacharbrackleft}{\kern0pt}g{\isacharcomma}{\kern0pt}\ m{\isacharcomma}{\kern0pt}\ p{\isacharcomma}{\kern0pt}\ h{\isacharbrackright}{\kern0pt}\ {\isasymturnstile}\ nid{\isacharprime}{\kern0pt}{\isacharprime}{\kern0pt}\ {\isasymleadsto}\ nid{\isacharprime}{\kern0pt}}}{\mbox{{\isacharbrackleft}{\kern0pt}g{\isacharcomma}{\kern0pt}\ m{\isacharcomma}{\kern0pt}\ p{\isacharcomma}{\kern0pt}\ h{\isacharbrackright}{\kern0pt}\ {\isasymturnstile}\ nid\ {\isasymleadsto}\ nid{\isacharprime}{\kern0pt}}}}
\end{center}%
\end{isamarkuptext}\isamarkuptrue%
}%EndSnip
\Snip{CanonicalizeIfNodeRules}{%
\begin{center}
\isa{\mbox{}\inferrule{\mbox{g{\isasymllangle}cond{\isasymrrangle}\ {\isacharequal}{\kern0pt}\ ConstantNode\ condv}\\\ \mbox{val{\isacharunderscore}{\kern0pt}to{\isacharunderscore}{\kern0pt}bool\ condv}}{\mbox{CanonicalizeIf\ g\ {\isacharparenleft}{\kern0pt}IfNode\ cond\ tb\ fb{\isacharparenright}{\kern0pt}\ {\isacharparenleft}{\kern0pt}RefNode\ tb{\isacharparenright}{\kern0pt}}}}\\[8px]
\isa{\mbox{}\inferrule{\mbox{g{\isasymllangle}cond{\isasymrrangle}\ {\isacharequal}{\kern0pt}\ ConstantNode\ condv}\\\ \mbox{{\isasymnot}\ val{\isacharunderscore}{\kern0pt}to{\isacharunderscore}{\kern0pt}bool\ condv}}{\mbox{CanonicalizeIf\ g\ {\isacharparenleft}{\kern0pt}IfNode\ cond\ tb\ fb{\isacharparenright}{\kern0pt}\ {\isacharparenleft}{\kern0pt}RefNode\ fb{\isacharparenright}{\kern0pt}}}}\\[8px]
\isa{\mbox{}\inferrule{\mbox{{\isasymnot}\ is{\isacharunderscore}{\kern0pt}ConstantNode\ g{\isasymllangle}cond{\isasymrrangle}}\\\ \mbox{tb\ {\isacharequal}{\kern0pt}\ fb}}{\mbox{CanonicalizeIf\ g\ {\isacharparenleft}{\kern0pt}IfNode\ cond\ tb\ fb{\isacharparenright}{\kern0pt}\ {\isacharparenleft}{\kern0pt}RefNode\ tb{\isacharparenright}{\kern0pt}}}}
\end{center}
}%EndSnip
\Snip{CanonicalizeIfNodeProof}{%
\begin{center}
\begin{isabelle}%
{\isasymlbrakk}g{\isasymllangle}nid{\isasymrrangle}\ {\isacharequal}{\kern0pt}\ before{\isacharsemicolon}{\kern0pt}\ CanonicalizeIf\ g\ before\ after{\isacharsemicolon}{\kern0pt}\isanewline
\isaindent{\ }g{\isacharprime}{\kern0pt}\ {\isacharequal}{\kern0pt}\ replace{\isacharunderscore}{\kern0pt}node\ nid\ after\ g{\isacharsemicolon}{\kern0pt}\ {\isacharbrackleft}{\kern0pt}g{\isacharcomma}{\kern0pt}\ p{\isacharbrackright}{\kern0pt}\ {\isasymturnstile}\ {\isacharparenleft}{\kern0pt}nid{\isacharcomma}{\kern0pt}\ m{\isacharcomma}{\kern0pt}\ h{\isacharparenright}{\kern0pt}\ {\isasymrightarrow}\ {\isacharparenleft}{\kern0pt}nid{\isacharprime}{\kern0pt}{\isacharcomma}{\kern0pt}\ m{\isacharcomma}{\kern0pt}\ h{\isacharparenright}{\kern0pt}{\isasymrbrakk}\isanewline
{\isasymLongrightarrow}\ nid\ {\isacharbar}{\kern0pt}\ g\ {\isasymsim}\ g{\isacharprime}{\kern0pt}%
\end{isabelle}
\end{center}
}%EndSnip
\Snip{example2}{%
\begin{isabelle}%
inputs\ensuremath{^{\mathit{g{\isasymllangle}nid{\isasymrrangle}}}}\ensuremath{_{\mathit{f}}}%
\end{isabelle}
}%EndSnip

\ifdefined\buildarxiv
\pagestyle{myheadings}
\fi

\title{A Formal Semantics of the GraalVM Intermediate Representation}
\titlerunning{A formal semantics of the GraalVM IR}

\author{Brae J. Webb\orcidID{0000-0003-3579-0244} \and
Mark Utting\orcidID{0000-0003-3134-6306} \and
Ian J. Hayes\orcidID{0000-0003-3649-392X}}

\authorrunning{B. J. Webb \and I. J. Hayes \and M. Utting}

\institute{The University of Queensland \email{\{B.Webb,M.Utting,Ian.Hayes\}@uq.edu.au}}

\begin{document}
\maketitle

\begin{abstract}

The optimization phase of a compiler is responsible for 
transforming an intermediate representation (IR) of a program into a more efficient form.
Modern optimizers, such as that used in the GraalVM compiler, 
use an IR consisting of a sophisticated graph data structure 
that combines data flow and control flow into the one structure.
As part of a wider project on the verification of optimization passes of GraalVM,
this paper describes a semantics for its IR within Isabelle/HOL.
The semantics consists of a big-step operational semantics for data nodes
(which are represented in a graph-based static single assignment (SSA) form)
and a small-step operational semantics for handling control flow
including heap-based reads and writes, exceptions, and method calls.
We have proved a suite of canonicalization optimizations and 
conditional elimination optimizations with respect to the semantics.
\end{abstract}

\section{Introduction}\labelsect{introduction}

Compilers are an essential ingredient of the computing base. 
Software developers need to be able to trust their compilers
because an error in a compiler can manifest as erroneous generated code
for any of the myriad of programs it compiles. 

This paper forms the first steps of a wider project that focuses on the verification 
of compiler optimization passes, a common source of compiler errors.
The project does not cover initial parsing, type checking and intermediate representation (IR) construction passes,
nor the final machine-dependent code generation pass.

The multi-pass structure of a compiler affords verification on a pass-by-pass basis.
An optimization pass transforms a program represented in the IR.
The verification of a pass involves proving that, for every IR input program,
the transformation implemented by the pass preserves the semantics of the program.
This task can be partitioned into:
\begin{itemize}
\item
defining a formal semantics for the IR,
\item
defining the optimizations as transformations of the IR, and
\item
verifying that the transformations are semantics preserving.
\end{itemize}

In this paper, we embark on the process of verifying the optimization passes
of an existing production compiler, GraalVM \cite{graal}, using Isabelle/HOL \cite{IsabelleHOL}.
We present a formalization of the IR used by the GraalVM compiler (Sect 3-6). %TODO: actual references
We briefly describe the validation of this semantics against the existing compiler implementation (\refsect{validation}),
then show the effectiveness of the semantics by proving two kinds of local optimizations (\refsect{optimization}).

The main contribution of this paper is to devise a formal semantics of the GraalVM IR in Isabelle/HOL \cite{IsabelleHOL}.
The IR combines control flow and data flow into a single `sea-of-nodes' graph structure \cite{click95},
rather than a more conventional control-flow graph with basic blocks representing sequential flow.
\refsect{GraalVM_IR} gives further details of the GraalVM Compiler.
As far as we are aware, this is the first formal semantics of a sea-of-nodes IR that covers method calls, with exceptions, as well as object reads and writes.
The semantics of the IR consists of the following components:
\begin{itemize}
\item
the graph representation corresponding to the GraalVM IR (\refsect{graph_model}),
\item
data-flow semantics that handles expression evaluation using a big-step operational semantics (\refsect{data_flow}),
\item
local control-flow semantics that handles control flow within a single method using a small-step operational semantics (\refsect{local_control_flow}),
\item
global control-flow semantics that handles method invocation and return, exceptions handling, and 
promotes the local control-flow semantics to a small-step operational semantics (\refsect{global_control_flow}).
\end{itemize}
Each stage builds on the previous.  Note that expression evaluation within the GraalVM IR is side-effect-free and terminating, so it is appropriate to use a big-step semantics that just returns the result, whereas for the control-flow semantics we use a small-step operational semantics to account for non-terminating programs and to accurately model the order of all side-effects, including object reads and writes, method calls, etc.

\section{GraalVM IR}\labelsect{GraalVM_IR}

The GraalVM Intermediate Representation (IR) is a sophisticated graph structure
that is designed to support implementation of efficient code optimizing transformations
(see \reffig{graph1} for an example graph).
A side-effect-free expression is represented by a data-flow subgraph that is acyclic (i.e.\ it is a DAG),
so that common subexpressions are represented by shared subgraphs 
(instead of by value numbering in traditional SSA form).
This has the advantage that data-flow dependencies are explicitly represented in the graph \cite{ProgramDependenceGraph_1987}.
Expressions with potentially observable side-effects, such as method invocations or field accesses,
are incorporated into the control-flow graph.

The IR combines both data-flow and control-flow aspects of a program within a single graph structure.
This graphical representation allows efficient implementation of optimizations equivalent to global value numbering and
global code motion optimization strategies \cite{Click1995}.

\stamp{
Additionally, almost all nodes include a Stamp field.
A stamp holds typing and range information of the value associated with the evaluation of the node.
The stamp information is determined by prior stages of static analysis and is used to perform some optimizations.
}
\begin{lstlisting}[language=Java,caption=A simplified \textsf{AddNode} class definition in GraalVM, float=t, label=nodeclass, frame=t]
class AddNode extends Node {
    @Input ValueNode x;
    @Input ValueNode y;
}
\end{lstlisting}

The GraalVM IR graph consists of many different kinds of nodes (over 200) with two main kinds of edges:
\begin{itemize}
\item \emph{input} edges that specify the data inputs of a node;
\item \emph{successor} edges that specify the control-flow successors of a node.
\end{itemize}

Nodes of the GraalVM IR are implemented in Java as a collection of Java classes which inherit from a base \textsf{Node} class. 
Each subclass of \textsf{Node} can specify their possible edge connections, either input or successor edges, by annotating fields 
that store references to other \textsf{Node} subclasses.
Listing \ref{nodeclass} shows a simplified example of one such \textsf{Node} subclass for an addition expression.
\textsf{AddNode} has two input edges $x$ and $y$ but it has no successors because it is a pure data-flow node.

\section{Graph Model in Isabelle/HOL}\labelsect{graph_model}

Our Isabelle/HOL model of the GraalVM IR graph has a close correspondence with the Java \textsf{Node} subclasses
but still supports efficient reasoning and pattern matching in Isabelle.
We use natural numbers%
\footnote{A more abstract representation would be better but using natural numbers allows us to utilise Isabelle code generation facilities.}
to identify nodes of the graph, 
and define an Isabelle datatype $IRNode$ (see \reffig{IRNode}) to model the concrete subclasses of the Java \textsf{Node} class.  
We developed a tool that uses Java reflection to traverse the GraalVM \textsf{Node} subclasses 
and generate the $IRNode$ datatype, 
including within each branch of the datatype the input edges, successor edges, and selected data attributes of each node, 
using the same names as in the Java source code 
but prefixed with ``$ir\_$'' to avoid name clashes (field names are global functions in Isabelle). 
We currently translate 45 of the concrete subclasses of the Java \textsf{Node} class into Isabelle, 
which corresponds to over 85\% of the nodes used to compile the Dacapo Java benchmark%
\footnote{\url{https://github.com/dacapobench/dacapobench}}
and is enough to handle simple example programs.
\stamp{
Typing information available through the stamp of a node is represented by a separate Stamp datatype (see \reffig{Stamp}). The stamp of a node is represented externally to the $IRNode$ datatype to reduce the complexity and assist in reasoning about optimizations.}
For the 60+ interface classes and abstract Java subclasses of \textsf{Node}, such as \textsf{BinaryArithmeticNode}, we also generate a corresponding Isabelle boolean function%
\footnote{In Isabelle/HOL ``$S \Rightarrow T$'' is the type of a function from $S$ to $T$.}
over the $IRNode$ type, such as:
\includesnippet{isbinary}

\begin{figure}
  \begin{small}
  \begin{displaymath}
  \begin{array}{l}
  \isacommand{type\_synonym}\ ID\ =\ nat\\
  \isacommand{type\_synonym}\ INPUT\ =\ ID\\
  \isacommand{type\_synonym}\ SUCC\ =\ ID\\[1ex]
  \isacommand{datatype}\ IRNode\ = \\
  \begin{array}{l}
    NoNode \\
    \Comment{Subclasses of FloatingNode} \\
    |\ ConstantNode (ir\_const:Value) \\
    |\ ParameterNode (ir\_index:nat) \\
    |\ ValuePhiNode (ir\_nid:ID) (ir\_values:INPUT\ list) (ir\_merge:INPUT)  \\
    |\ NegateNode (ir\_value:INPUT) \\
    |\ AddNode (ir\_x:INPUT) (ir\_y:INPUT) \\
    |\ MulNode (ir\_x:INPUT) (ir\_y:INPUT) \\
    |\ IntegerLessThanNode (ir\_x:INPUT) (ir\_y:INPUT) \\
    |\ . . . \\
    \Comment{Control flow (fixed) nodes} \\
    |\ StartNode\ ...\ (ir\_next:SUCC) \\
    |\ IfNode (ir\_condition:INPUT) \\
    \qquad (ir\_trueSuccessor:SUCC) (ir\_falseSuccessor:SUCC) \\
    |\ BeginNode (ir\_next:SUCC) \\
    |\ EndNode \\
    |\ LoopBeginNode (ir\_ends:INPUT\ list)\ ...\ ...\ (ir\_next:SUCC) \\
    |\ LoopEndNode (ir\_loopBegin:INPUT) \\
    |\ LoopExitNode (ir\_loopBegin:INPUT)\ ...\ (ir\_next:SUCC) \\
    |\ MergeNode  (ir\_ends:INPUT\ list)\ ...\ (ir\_next:SUCC) \\
    |\ NewInstanceNode (ir\_nid:ID) (ir\_instanceClass:string) \ ...\ (ir\_next:SUCC) \\
    |\ LoadFieldNode (ir\_nid:ID) (ir\_field:string) (ir\_object\_opt:INPUT\ option) \\
    \qquad (ir\_next:SUCC) \\
    |\ StoreFieldNode (ir\_nid:ID) (ir\_field:string) (ir\_value:INPUT) \ ...\ \\
    \qquad (ir\_object\_opt:INPUT\ option) (ir\_next:SUCC) \\
    |\ RefNode(ir\_next:SUCC) \\
    \Comment{Interprocedural nodes} \\
    |\ ReturnNode (ir\_result\_opt:INPUT\ option)\ ...\ \\
    |\ InvokeNode (ir\_nid:ID) (ir\_callTarget:INPUT)\ ...\ ...\ ...\ (ir\_next:SUCC) \\
    |\ InvokeWithExceptionNode (ir\_nid:ID) (ir\_callTarget:INPUT)\ ...\ ...\ ...\ \\
    \qquad (ir\_next:SUCC) (ir\_exceptionEdge:SUCC) \\
    |\ MethodCallTargetNode (ir\_targetMethod:string) (ir\_arguments:INPUT\ list) \\
    |\ UnwindNode (ir\_exception:INPUT)\\
  \end{array}
  \end{array}
  \end{displaymath}
  \end{small}
  \caption{An extract of the Isabelle datatype definition of the IR graph nodes (some node types and fields are omitted or abbreviated to save space).}
  \labelfig{IRNode}
  \end{figure}

\reffig{IRNode} gives the Isabelle representation of the graph nodes.%
\footnote{All theories are available at \url{https://github.com/uqcyber/veriopt-releases/tree/atva2021}.}
$ConstantNode$ corresponds to a Java constant, so has a value constant as its only field, with no input or successor edges.
Similarly, $ParameterNode$ has a single natural number field that is an index into the list of parameter values of the current method.
Binary expression nodes (like $AddNode$) have two input expression edges, named $ir\_x$ and $ir\_y$.
The data-flow aspects of merging multiple control-flow paths are handled by a $\phi$-node (abbreviating $ValuePhiNode$) for each value that is dependent 
on the path used to reach an associated merge node (e.g. $MergeNode$).
The semantics of $\phi$-nodes
is explained more fully in \refsect{local_control_flow}, but note that a $\phi$-node
has a pseudo-input edge called $ir\_merge$ that references the merge node associated with the $\phi$-node,
and a list of input edges $ir\_values$ that is in one-to-one correspondence with the control-flow edges into that merge node.
To illustrate how the structure of a node influences its semantics, consider an $IfNode$.
An $IfNode$ has one input edge for its boolean condition, and two successor edges, one to take when the condition evaluates to true and the other successor edge to take when it evaluates to false. 

\stamp{
\begin{figure}
\includesnippet{Stamp}
\caption{Isabelle/HOL representation of Stamp information from GraalVM}
\labelfig{Stamp}
\end{figure}
}

In addition to explicit (named) input and successor fields, the Java \textsf{Node} classes use annotations and meta-data in each subclass to provide \emph{generic} access functions for accessing the list of all inputs of an arbitrary subclass, and similarly for all successors.  Such generic access is helpful for implementing visitor patterns that walk the graph, etc.  In Isabelle, we provide the equivalent functionality by defining two functions over $IRNode$, $inputs\text{-}of$ and $successors\text{-}of$, in the following style,
in which ``$\cdot$'' represents list cons.
\begin{isabelle}
\includesnippet{inputs_of}
\end{isabelle}

\indent We model an IR graph for a single method as a partial map ($\isasymrightharpoonup$) from node $ID$s to $IRNode$s 
\stamp{and $Stamp$s }with a finite domain.
\begin{isabelle}
\includesnippet{graphdefnostamp}
\end{isabelle}
A finite domain is a requisite property for code generation used by validation efforts (see \refsect{validation}), 
however, we have found reasoning to be more straightforward with total functions and hence we introduce the kind function, 
denoted $g\langle\langle nid \rangle\rangle$, 
that is a total function that performs lookup on the underlying partial function, $g$, resulting in $NoNode$ for identifiers with no mapping.
In addition, we lift the domain function to the function $ids$ and introduce functions $inputs$, $succ$, $usages$, and $predecessors$ that, given a graph and a node identifier,
produce the sets of input, successor, usage, and predecessor node ids, respectively.

There are several conditions that a graph $g$ should satisfy to be well-formed, such as being closed, 
i.e.\ all inputs and successors identify nodes within the graph (that is, within $ids~g$).  The key invariants that we have needed so far are shown in \reffig{graphinvar}, and include the edge-closure properties, as well as the requirement that node zero should be the $StartNode$ for the method represented by the graph, and that all the nodes in the graph are proper nodes, rather than $NoNode$.
Additionally, end nodes need to have at least one usage which is treated as the pseudo-successor edge for an end node. The input edges of a merge node are used by $\phi$ nodes to determine the value for a $\phi$ node, the number of input edges of any $\phi$ node must match the number of input edges of its associated merge node to ensure correct execution.
We expect to add further invariants in the future as we prove deeper properties of the graph.  Indeed, one of the expected benefits of this project is to discover important IR invariants that are currently implicit in the way that the GraalVM compiler constructs and uses the graph, and to:
\begin{itemize}
\item prove that those invariants are preserved by the graph transformations that represent optimizations;
\item document those invariants explicitly and implement them in the Java code base so that they can be checked at runtime during test runs of the compiler.
\end{itemize}

\begin{figure}
  \begin{minipage}[t]{.545\linewidth}
    \includesnippet{wf_start_def}
  \end{minipage}
  \begin{minipage}[t]{.545\linewidth}
    \includesnippet{wf_ends_def}
  \end{minipage}
  \begin{minipage}[b]{.545\linewidth}
    \includesnippet{wf_phis_def}
  \end{minipage}
  \begin{minipage}[b]{.545\linewidth}
    \includesnippet{wf_closed_def}
  \end{minipage}
  
\includesnippet{wf_graph_def}\vspace*{-1ex}
\caption{Isabelle well-formedness graph invariants.}
\labelfig{graphinvar}
\end{figure}

An $IRGraph$ represents a single method.
In the GraalVM compiler, to uniquely identify a method, one needs not only its name but the class in which it is defined
and the types of its parameters to handle method overloading (as in Java \cite{jvm-spec}).
Together these form the method's signature, which is represented by the type $Signature$.
Programs are represented as a partial function from method signatures to their $IRGraph$.
\includesnippet{programdef}

\section{Data-flow semantics}\labelsect{data_flow}

In a programming language like Java, expression evaluation may involve side effects, such as calling a method.
The GraalVM, and hence our semantics, treats nodes that may have a side effect differently.
These nodes are included in the control-flow graph so that they are evaluated as part of the control-flow semantics 
(see \refsect{local_control_flow})
and hence the order of their evaluation is preserved.
When one of these nodes (with node identifier $nid$, say) is evaluated as part of the control flow semantics,
the calculated value is saved under the node identifier $nid$ in a mapping $m$ from node identifiers to values,
which we refer to as the \emph{method state}.

The data-flow semantics handles the evaluation of side-effect-free expressions,
which are represented by a directed acyclic (sub-)graph (DAG),
in which internal nodes are operators (with input arguments that are graph node ids)
and leaf nodes are either constant nodes, parameter nodes, or control-flow nodes representing expressions that may have had side effects,
e.g.\ a method invocation node.
These control-flow nodes have their current value stored in the method state $m$ under their node identifier, 
with $m~nid$ giving the current value associated with (leaf) node $nid$.
The values of the parameters are given by a list of values $p$, with $p_{[i]}$ giving the value of the $i^{th}$ parameter.

\begin{figure}[ht]
    \caption{Data-flow semantics for a subset of nodes}
    \labelfig{data_semantics}
    \includesnippet{ExpressionSemantics}    
\end{figure}
For a graph $g$, method state $m$, and list of parameter values $p$, 
in our big-step operational semantics for expressions, 
an evaluation of a node $n$ to a value $v$ is represented as
\[
   [g,\ m,\ p]\ {\isasymturnstile}\ n\ {\isasymmapsto}\ v.
\]
A sample of the 27 evaluation rules for data nodes is given in \reffig{data_semantics}.
Note that for the $AddNode$, the $+$ is overloaded to use the Isabelle/HOL WORD library to add two fixed-size integers, 
so that integer arithmetic follows Java's twos-complement semantics with wrapping upon overflow.

Each parameter node contains the index $i$ of its parameter in the formal parameter list, with $p_{[i]}$ giving the parameter's value.
Control-flow nodes for expressions with side effects (such as $ValuePhiNode$, $InvokeNode$, $NewInstanceNode$, $LoadFieldNode$)
extract the current value of the node from the method state $m$.
Each of these node types also has a rule in the control-flow semantics that triggers their evaluation and updates $m$ with the result
(see \refsect{local_control_flow}).
The control-flow semantics requires the ability to evaluate a list of expressions, $nids$, to a list of values, $vs$, written, 
\[
  [g,~m,~p]~\vdash~nids~{\isasymlongmapsto}~vs,
\] 
(note the longer arrow), which is the obvious lifting of evaluation of a single expression to evaluate each expression in the list
(not detailed for space reasons).

\section{Local control-flow semantics}\labelsect{local_control_flow}

To support object orientation, the semantics requires a heap to store objects. 
We define a heap in the form of a function $h$ that for an object reference $r$ and a field name $f$ gives the value of that field for that object, $h~r~f$ \cite{heap-reps-2011}.
Note that while the heap is always finite, its size is unbounded.
\reffig{heapdef} defines our heap representation. 
$Heap$ is a type that maps object references and field names to values. 
$DynamicHeap$ expands $Heap$ to track the next free object reference%
\footnote{The operation for allocating a new object could nondeterministically choose any unused object reference,
but we have made it a deterministic function that allocates the next location to facilitate the use of Isabelle code generation facilities.
}
in the heap, $Free$, 
each time a new object is instantiated the next free object reference is incremented and the current free object reference is used.
The supporting definitions, \textit{h-load-field}, \textit{h-store-field}, and \textit{h-new-inst}, 
are used by the semantics of the load (\ref{step:load}) and store (\ref{step:store}) field nodes in \reffig{control_semantics}.

\begin{figure}
\begin{isabelle}
\includesnippet{heapdef}
\end{isabelle}
\caption{Isabelle model of a heap and supporting definitions}
\labelfig{heapdef}
\end{figure}

The control-flow semantics local to a method is given by a small-step operational semantics.
A configuration consists of a triple of the current node id, $nid$, the method state, $m$, 
as used in expression evaluation, and the heap, $h$.
The transition
\[
  [g,\ p]\ {\isasymturnstile}\ {\isacharparenleft}nid{\isacharcomma}\ m{\isacharcomma}\ h{\isacharparenright}\ {\isasymrightarrow}\ {\isacharparenleft}nid{\isacharprime} {\isacharcomma}\ m{\isacharprime}{\isacharcomma}\ h{\isacharprime}{\isacharparenright}, 
\]
can be read as, within the context of graph, $g$, and list of parameter values, $p$,
an execution step can transition from configuration $(nid, m, h)$ to configuration $(nid', m', h')$.
The node id acts as a program counter for the graph representation of the method.
For a configuration, $(nid,m,h)$, to be valid, $nid$ must be a control flow node within $g$,
$p$ must give values for all parameters to the current method,
and
$m$ gives the values for all control-flow nodes that represent expressions with side effects 
that have been reached in the current invocation of the method.

\begin{figure}[ht]
    \caption{Control Node Semantics}
    \labelfig{control_semantics}
    \includesnippet{StepSemantics}
\end{figure}

\reffig{control_semantics} shows most of the rules for the local control-flow semantics --- to save space we omit the load and store rules for static fields, where the object pointer is $None$ rather than $(Some\;obj)$. 
A number of nodes have a control-flow behaviour of a no-op; we group them together as sequential nodes.
Their semantics (\ref{step:seq}) is a transition from the current node to the node attached to the only successor edge. 
An $IfNode$ (\ref{step:if}) chooses to transition to the first ($tb$) or second ($fb$) successor edge
based on the evaluation of the condition expression.

\newsavebox{\factcode}

\newcommand{\cfbox}[2]{%
    \colorlet{currentcolor}{.}%
    {\color{#1}%
    \fbox{\color{currentcolor}#2}}%
}

\newsavebox{\mybox}
\newenvironment{Notes}
{\begin{lrbox}{\mybox}}
{\end{lrbox}\fbox{\usebox{\mybox}}}

\begin{lrbox}{\factcode}
\begin{lstlisting}[language=Java,frame=t,linewidth=0.28\textwidth,xleftmargin=0pt,framexleftmargin=5pt,escapeinside={(*@}{@*)}]
int fact(int n) {
  int result = 1;
  while (n > 1) {
    result *= n;
    n = n - 1;
  }
  return result;
}
\end{lstlisting}
\end{lrbox}

\begin{figure}
  \floatsetup{valign=t, heightadjust=all, capposition=bottom}
  \ffigbox{%
    \begin{subfloatrow}[2]
      \ffigbox{\vfill\usebox{\factcode}\vfill}{\caption{Java factorial program \label{fact}}}
      \ffigbox{\includegraphics[width=0.5\textwidth]{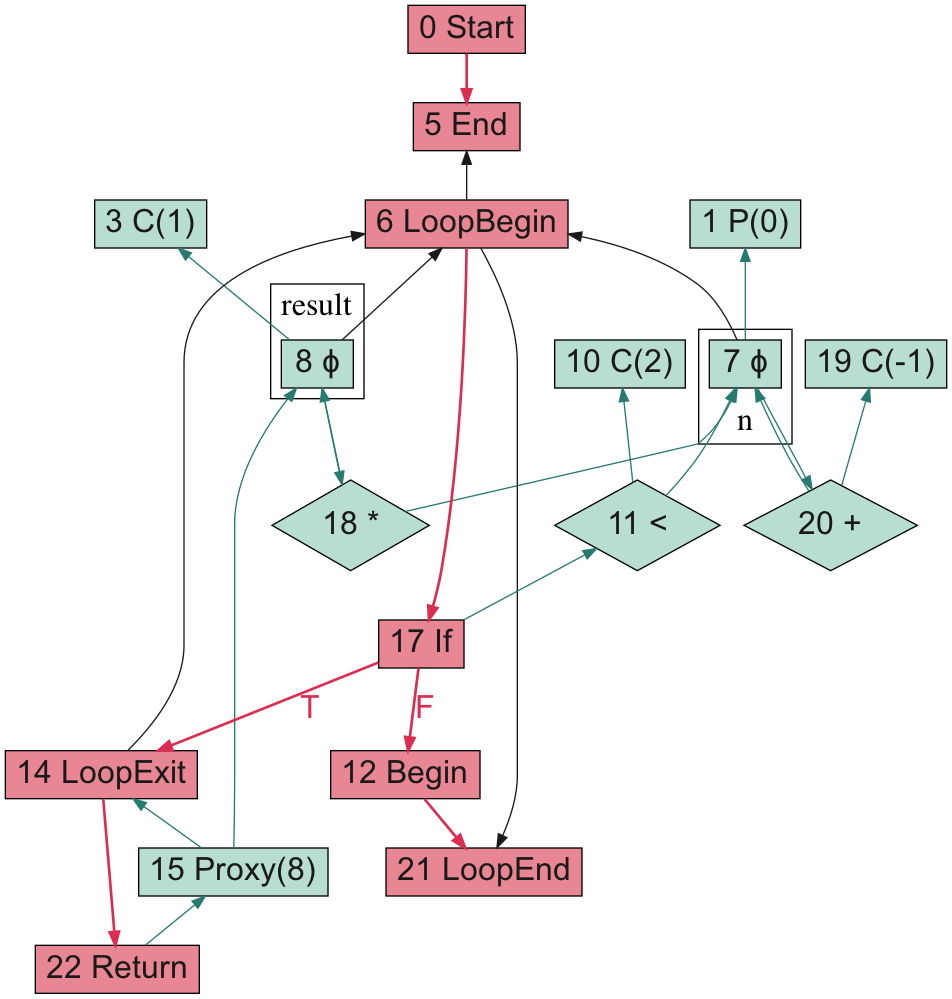}}{\caption{Factorial program IR graph} \labelfig{graph1}}
    \end{subfloatrow}}
    {\caption{Example factorial program transformed into a GraalVM IR graph}}
\end{figure}

Our approach to handling $\phi$ nodes is similar to that used by Demange \textit{et al.}\ for their formalization of 
reasoning about the sea of nodes in Coq \cite{sea-of-nodes-semantics}.
End nodes (\ref{step:end}) represent the end of a basic block in SSA terminology.
Each end node forms an input to a merge node
and each merge node has an associated set of $\phi$ nodes, 
each of which represents a value that is dependent 
on which path was taken to reach the end node, and hence the merge node.
When an end node is reached, the method state $m$ of each associated $\phi$ node is updated 
with the value of its associated expression DAG in the current state, $m$.
This process is best explained via the example in \reffig{graph1}, in which 
nodes 3, 10 and 19 are constant nodes,
node 20 is an \textsf{AddNode},
node 18 is a \textsf{MulNode},
node 11 is a \textsf{IntegerLessThanNode},
$\phi$-node 8 represents the value of the local variable \textsf{result},
and 
node 1 corresponds to the parameter \textsf{n},
which provides the initial value of $\phi$-node 7,
which represents the variable \textsf{n} within the loop.
The \textsf{ProxyNode} 15 is the value of the $\phi$-node 8 (i.e.\ \textsf{result}) but has an additional dependency on the \textsf{LoopExitNode} 14
to ensure the value is that after loop exit.
Note that the value of \textsf{AddNode} 20 is calculated using the inputs constant -1 and the $\phi$-node 7,
representing the previous value of \textsf{n}, to give the new value of the $\phi$-node 7
(hence the double-headed arrow between nodes 7 and 20). 
Given
\begin{itemize}
  \item $merge$, the id of the merge node \textsf{LoopBegin} \noderef{6},
  \item usage $\phi$ nodes of $merge$, $phis$ = [$\phi_1$ \noderef{7}, $\phi_2$ \noderef{8}\,]
  \item input end nodes of $merge$, $ends$ = [\textsf{End} \noderef{5}, \textsf{LoopEnd} \noderef{21}\,]
  \item inputs of $\phi_1$ \noderef{7} excluding $merge$,  [\textsf{ParameterNode} P(0) \noderef{1}, \textsf{AddNode} + \noderef{20}\,]
  \item inputs of $\phi_2$ \noderef{8} excl.\ $merge$, [\textsf{ConstantNode} C(1) \noderef{3}, \textsf{MultiplyNode} * \noderef{18}]
\end{itemize}
when
\begin{itemize}
  \item $End$ \noderef{5} is reached
  \begin{enumerate}
    \item evaluate the first input edge to all $phis$
    in the original method state, $m$, i.e. for  \noderef{1}, $[g,m,p] \vdash P(0) \mapsto r_1$ and
    for \noderef{3} $[g,m,p] \vdash C(1) \mapsto 1$.
    \item update $m$ to map the values of the evaluated expressions to each $\phi$ node, i.e. $m'(\phi_1) = r_1$ and $m'(\phi_2) = 1$.
  \end{enumerate}
  \item $LoopEnd$ \noderef{21} is reached
  \begin{enumerate}
    \item evaluate the second input edge to all $phis$
    in the original method state, $m$, 
    i.e. for \noderef{20} $[g,m,p] \vdash AddNode(\noderef{7},\noderef{19}) \mapsto r_1$ and
    for \noderef{18} $[g,m,p] \vdash MulNode(\noderef{8}, \noderef{7}) \mapsto r_2$.
    Note that when the evaluation reaches a $\phi$ node, it refers to the (previous) value of the $\phi$ node in $m$, i.e. $m(\phi)$.
    \item update $m$ to map the values of the evaluated expressions to each $\phi$ node, i.e. $m'(\phi_1) = r_1$ and $m'(\phi_2) = r_2$.
  \end{enumerate}
\end{itemize}

More generally, a merge node may have a list of input end nodes, $ns$,
and any number of associated $\phi$ nodes,
each of which has a list of input expressions, each of which is of the same length as $ns$.
When the merge node is reached via its $i^{th}$ input end node,
the value of each associated $\phi$ node is updated within $m$
to the value of the $(i+1)^{th}$ input expression of the $\phi$ node using method state $m$
(the $i+1$ offset is because input edge zero of a $\phi$ node is used to connect to its merge node).

When a $NewInstanceNode$ is reached in the control flow (\ref{step:newinst}), 
space is allocated in the heap for a new object $ref$ using the function \textit{h-new-inst} function (\reffig{heapdef}).
The value associated with the $NewInstanceNode$ is updated in $m'$ to the new object reference $ref$
so that subsequent data-flow evaluations of the $NewInstanceNode$ evaluate to $ref$. 

A $LoadFieldNode$ (\ref{step:load}) contains
a field name $f$ and an optional input edge to a node that must evaluate to an object reference, $obj$. 
The \textit{h-load-field} function (\reffig{heapdef}) reads the value from the heap based on the object reference and field name.
The resulting value, $v$, is then stored in $m'$ under the node id of $LoadFieldNode$ 
so that subsequent data-flow evaluations of the $LoadFieldNode$
result in $v$.

Similar to the $LoadFieldNode$, the $StoreFieldNode$ (\ref{step:store}) contains a field identifier, $f$, and an optional
input edge to a node which must evaluate to an object reference, $obj$. A $StoreFieldNode$ also has an input edge to a node, $newval$, that
is evaluated to a value, $val$ and stored in the heap.
The \textit{h-store-field} function (\reffig{heapdef}) stores $val$ in the updated heap, $h'$,
corresponding to the field $f$ and object reference, $obj$.
Note that null pointer dereferences are checked by a separate (dominating) $GuardNode$ (not covered in this paper) and hence null pointer dereferences are not an issue for load and store field.
To save space, we omit load and store for static fields --- these do not evaluate an object reference.

\section{Global control-flow semantics}\labelsect{global_control_flow}

The semantics in \refsect{local_control_flow} only handles control flow within a single method.
To handle method calls and returns, we lift the semantics to a richer global configuration 
that consists of a pair, $(stk,h)$, containing a stack, $stk$, of local configurations for each called but not yet returned method
and a global heap, $h$.
The stack contains tuples of the form $(g,nid,m,p)$, in which $g$ represents the method's graph,
$nid$ is a node id (the program counter) within $g$, $m$ is the method state, and $p$ is the list of of parameter values, 
as for the data-flow semantics.
The $IRGraph$ of the method with signature $s$ in program $P$ (of type $Program$) is given by $P~s$.

\reffig{interprocedural_semantics} gives a small-step semantics for global control flow.
Given a program $P$, a transition of the form
$
  P \vdash (stk, h) \longrightarrow (stk', h')
$
represents a step from a configuration stack $stk$ and heap $h$ to a new stack $stk'$ and heap $h'$. 
Stacks are represented as lists, so $(g,nid,m,p) {\isasymcdot} stk$ represents 
a stack with top as the local configuration $(g,nid,m,p)$ and remainder of the stack, $stk$.

\begin{figure}[ht]
    \caption{Interprocedural Semantics}
    \labelfig{interprocedural_semantics}
    \includesnippet{TopStepSemantics}
\end{figure}

Local control-flow transitions are promoted to global control-flow transitions in which the top of stack is updated according to the local transition step (\ref{top:lift}).

For an $InvokeNode$ (\ref{top:invoke}), its list of actual parameter expressions, $arguments$, is evaluated 
to give the list of parameter values, $p'$.
The method state $m'$ for the invoked method is initially empty ($new{\isacharunderscore}{\kern0pt}map{\isacharunderscore}{\kern0pt}state$).
The method being invoked is determined by the $MethodCallTargetNode$, 
which is attached via an input edge to an $InvokeNode$. 
The $MethodCallTargetNode$ contains the signature, \textit{targetMethod}, of the invoked method.
A new local configuration consisting of 
the graph of the invoked method, $targetGraph$, 
a method start node id of zero, 
the method state $m'$,
and the list of parameter values $p'$
is pushed onto the stack.

For a $ReturnNode$ (\ref{top:return}), the return expression is optional.
Here we only consider the case in which there is some return expression.
The return value, $v$, is calculated using the top-of-stack graph $g$, method state $m$ and parameters $p$ (i.e.\ the called method).
The second stack element is a local configuration containing the graph of the calling method, \textit{cg}, 
id of the invocation node, $cnid$, the method state at the point of call, $cm$,
and the parameters of the calling method, $cp$.
The top two elements of the stack are replaced by 
a single local configuration consisting of the calling method's graph $cg$,
the successor $cnid'$ of invocation node $cnid$,
a new method state $cm'$ that updates $cm$ to map
the invocation node $cnid$ to the returned value, $v$,
and the parameters to the calling method, $cp$.

Certain methods can result in exceptions rather than regular returned values.
Calls to these methods are made using the $InvokeWithExceptionNode$. The invocation of these methods is handled with the same semantics as $InvokeNode$.
An $UnwindNode$ (\ref{top:unwind}) indicates that 
an exception has been thrown.
The control-flow path when an $UnwindNode$ is reached is determined by the \textit{exEdge} successor of the calling $InvokeWithExceptionNode$.
The $InvokeWithExceptionNode$ is the node on the second top of the stack when an $UnwindNode$ is reached.
The top two elements of the stack are replaced by a single local configuration consisting of the graph of the calling method, $cg$, 
the \textit{exEdge} successor of the $InvokeWithExceptionNode$, and
the method state $cm$ updated so that the $InvokeWithExceptionNode$ maps to the object reference $e$ of the exception that was thrown.

\section{Validation of Execution Semantics} \labelsect{validation}

The GraalVM compiler contains thousands of unit test cases, and many of these define a standalone method.
Each test checks that its unoptimized and optimized execution give the same result.
We have added code to intercept such tests and translate the unoptimized IR graph, the input parameter values, and the expected result into our Isabelle IR graph notation.
We can then use Isabelle's code generation mechanism to execute the Isabelle IR graph of the method with the given input parameters, 
and check if the result matches.

We have translated and executed over 1400 of these unit tests so far, and after fixing a minor boolean-to-integer conversion issue and adding support for initializing static fields before the method is called, they all return the expected result.  This gives us some initial confidence that our execution semantics corresponds to the GraalVM IR semantics.  Any remaining differences will become apparent during the correctness proofs of optimizations.

\section{Proving Optimizations}\labelsect{optimization}

The GraalVM compiler contains a comprehensive canonicalization phase.
Subsequent optimization phases rely on the canonicalization phase to minimize the forms which an IR can take.
The majority of the canonicalization optimizations do not rely on additional static analysis processes, so 
are good case studies for the process of proving local optimizations.
A canonicalization of a data-flow node within a graph $g_1$, 
replaces a data-flow node in $g_1$ at $nid$ with a new node and
may introduce additional new nodes with fresh node ids to form a new graph $g_2$.
The replacement must maintain the property that the subgraph is acyclic.
While the new node at $nid$ may no longer reference some node ids that the original node at that position did,
the unreferenced nodes are left in the graph because there may be other references to those nodes elsewhere in graph.
To show the correctness of these forms of canonicalization optimizations,
noting that expression evaluation has been shown to be deterministic,
it is sufficient to show that for all method states $m$, 
evaluating the new node at $nid$ gives the same value as evaluating the old node at $nid$, i.e.
\[
 \forall m,~p ~.~ ([g_1,~m,~p]~\vdash g_1 \isasymllangle nid \isasymrrangle \mapsto v) \longleftrightarrow ([g_2,~m,~p]~\vdash g_2 \isasymllangle nid \isasymrrangle \mapsto v).
\]
For example, we have completed proofs of correctness of optimizations of conditional expressions (Java's \textsf{(c ? v1 : v2)}).

As an example of a canonicalization of the control-flow graph, 
we define a set of optimizations for the $IfNode$ in \reffig{canonical_if_rules}.
We show the optimization where an $IfNode$ with a constant condition is replaced by a $RefNode$ to either the true or false branch, where a $RefNode$ is a sequential node that just transitions to its successor.
In addition, we give the optimization where both successor edges of the $IfNode$ are equal, replacing with a $RefNode$ to one of the (equal) branches.
Note that these optimizations bypass the condition evaluation but as that is side effect free, it is of no consequence.
\begin{figure}[ht]
  \includesnippet{CanonicalizeIfNodeRules}
  \caption{Canonicalization rules for an $IfNode$}\labelfig{canonical_if_rules}
  \end{figure}

We prove that the canonicalization rules 
are correct by showing that, given:
\begin{itemize}
  \item a node, $before$, where $g\langle\langle nid\rangle\rangle$ = $before$;
  \item that $before$ can be canonicalized to the node $after$;
  \item a graph, $g'$, where the node at $nid$ has been replaced by $after$;
\end{itemize}\vspace{-0.8ex}
then we can prove that $g'$ has the same behaviour as $g$ starting from node $nid$ in both graphs.

Thus far, we have encoded and proved exploratory components of the canonicalization phase and the entirety of the conditional elimination phase allowed by our subset of nodes.
The techniques used for the requisite static analysis during the conditional elimination phase are to be the subject of future papers.

\section{Related Work}\labelsect{RelatedWorks}

The closest research to that presented here is the work of Demange \textit{et al.} \cite{sea-of-nodes-semantics} who provide the semantics of an abstract sea-of-nodes representation in Coq, which focuses on the semantics of $\phi$ nodes and regions. The semantics is used to prove a semantic property and a simple optimization transformation. Their formalization allows properties of the abstract sea-of-nodes representation to be proven in isolation. We offer a variant of this semantics that matches the concrete implementation of a production compiler, and we extend the approach to handle interprocedural calls and a heap-based object model.

Two notable verified compiler projects are CompCert \cite{compcert}, for a subset of C verified in Coq, and CakeML \cite{cakeml}, for a subset of ML verified in HOL4. These are both substantial projects verifying end-to-end correctness of their respective compilers from source code to generated machine code. Unlike these projects, this project targets only the optimization phase of the compiler, a common source of issues, rather than full end-to-end verification.

JinjaThreads \cite{jinjathreads} is a substantial formalization effort of the Java language semantics in Isabelle/HOL. Unlike our project, JinjaThreads focuses on directly formalizing the language semantics, rather than a language-agnostic IR. As the GraalVM IR is implemented in Java, one plausible approach to our project would be to use the JinjaThreads formalization to prove optimizations correct.  However, such proofs would have been undoubtedly laborious, so we have instead chosen to introduce a semantics to capture the IR semantics directly and allow optimizations to be more easily expressed and proved.

VeLLVM \cite{vellvm} formalizes the LLVM \cite{llvm} IR semantics using the Coq proof assistant. While the approach is similar, the target IR is substantially different. LLVM shares some common properties such as being in SSA form, but the GraalVM IR is a sea-of-nodes graph structure that unifies a program's control-flow and data-flow, while the LLVM IR is in traditional basic block SSA form. 

K-LLVM \cite{kllvm} is another formalization effort for the LLVM IR that does not directly expand on VeLLVM but expands the formalized feature set by offering a separate formalization implemented in $\mathbb{K}$. $\mathbb{K}$ is a framework designed for formalizing language semantics, which can produce language interpreters as well as export to Isabelle/HOL to allow proofs based on the specification.

\section{Conclusions} \labelsect{conclusions}

We have described an Isabelle model and execution semantics for the sophisticated sea-of-nodes graph structure \cite{click95} that is used as the internal representation in the GraalVM optimizing compiler \cite{Graal-IR-2013-10}. Additionally, we have proved several suites of local optimizations correct according to the semantics. 

In future work, we plan to tackle more global optimizations that transform the input graph in more complex ways.
In the longer term, we also want to explore expressing optimizations in a high-level notation that can more easily be transformed into Isabelle (for correctness proof purposes) as well as into Java code that implements the graph transformation, in order to have a tight connection between the Java and Isabelle graph transformations.

\paragraph*{Acknowledgements}
Mark Utting's position and Brae Webb's scholarship are both funded in part by a gift from Oracle Labs.
Thanks especially to Cristina Cifuentes, Paddy Krishnan and Andrew Craik from Oracle Labs Brisbane for their helpful feedback, and to the Oracle GraalVM compiler team for answering questions.  Thanks to Chris Seaton for helping us extend the SeaFoam IR visualization tool to output the graph in Isabelle syntax.
Thanks also to Kristian Thomassen 
for his work on the semantics of $\phi$-nodes and
Sadra Bayat Tork 
who investigated IR graph invariants in the GraalVM compiler.

\bibliographystyle{splncs04}
\bibliography{references}

\end{document}